\documentclass[reprint,twocolumn,superscriptaddress,longbibliography,footinbib]{revtex4-2}

\usepackage{graphicx}
\usepackage{amsmath}
\usepackage{amssymb}
\usepackage{mathtools}
\usepackage{braket}
\usepackage{multirow}
\usepackage{hyperref}
\usepackage{cleveref}
\usepackage{makecell}
\usepackage{verbatim}

\DeclareGraphicsExtensions{.png,.jpg,.eps}
\usepackage{xcolor}
\hypersetup{
    colorlinks,
    linkcolor={red!50!black},
    citecolor={blue!80!black},
    urlcolor={blue!80!black}
}

\renewcommand{\bm}{\boldsymbol}

\def\bk{{\boldsymbol{k}}}

\def\m1{{^{-1}}}

\begin{document}

%%%%%%%%%%%%%%%%%%%%%%%%%%%%%%%%%%%%%%%%%

\author{Bastian Zinkl}
\affiliation{Institute for Theoretical Physics, ETH Zurich, CH-8093 Zurich, Switzerland}

\author{Aline Ramires}
\affiliation{Condensed Matter Theory Group, Paul Scherrer Institute, CH-5232 Villigen PSI, Switzerland}

\title{Sensitivity of superconducting states to the impurity location in layered materials 
}

%%%%%%%%%%%%%%%%%%%%%%%%%%%%%%%%%%%%%%%%%
\begin{abstract}
The family of multi-layered superconductors derived from the doped topological insulator Bi$_2$Se$_3$ has been found to be unusually robust against non-magnetic disorder. Recent experimental studies have highlighted the fact that the location of impurities could play a critical role for this puzzling robustness. Here we investigate the effects of four different types of impurities, on-site, interstitial, intercalated and polar, on the superconducting critical temperature. We find that different components of the scattering potential are active depending on the impurity configuration and choice of orbitals for the effective low-energy description of the normal state.  For the specific case of Bi$_2$Se$_3$-based superconductors, we find that only the symmetric share of impurity configurations contribute to scattering, such that polar impurities are completely inactive. We also find that a more dominant mass-imbalance term in the normal-state Hamiltonian can make the superconducting state more robust to intercalated impurities, in contrast to the case of on-site or interstitial impurities. 

\end{abstract}
%%%%%%%%%%%%%%%%%%%%%%%%%%%%%%%%%%%%%%%%%

\date{\today}
\maketitle

%%%%%%%%%%%%%%%%%%%%%%%%%%%%%%%%%%%%%%%%%
\section{Introduction}
%Naive expectations on the effects of impurities in simple SCs

The effects of impurities in simple superconductors are well understood in terms of elementary symmetry arguments, elegantly summarized in what is known as Anderson's theorem \cite{Anderson:1959, Anderson:1984}. In fully-gapped conventional superconductors, pairs of electrons are formed related by time-reversal symmetry. As a consequence, impurities have a detrimental effect to superconductivity only if they break this key symmetry, namely, if the impurities are magnetic. For unconventional nodal superconductors both magnetic and non-magnetic disorder have a negative effect on the critical temperature \cite{Abrikosov:1959, Suhl:1959}. The sensitivity of superconductors to non-magnetic disorder has therefore been taken as a strong indication of the unconventional nature of the order parameter, as observed in UPt$_3$ \cite{Dalichaouch:1995}, Sr$_2$RuO$_4$ \cite{Mackenzie:1998}, and in the cuprates \cite{Fujita:2005}, to name a few.

%Complex superconductors
In contrast, the effects of impurities in complex superconductors have only recently started to attract attention, mostly motivated by the phenomenology of iron-based superconductors \cite{Senga:2008,Onari:2009,Wang:2013}. Complex superconductors are characterized by multiple Fermi surfaces, or by single Fermi surfaces emerging from a combination of multiple internal degrees of freedom. In this context, one result that goes beyond the prediction of Anderson's theorem concerns fully-gapped superconductors in multi-band systems. A two-band system with full gaps of opposite signs is known to be sensitive to non-magnetic disorder \cite{Senga:2008,Onari:2009,Wang:2013}. Naturally, extra sensitivity to disorder of such fully-gaped systems is in principle not a desirable feature to explore potential applications of these materials.

%Unexpected robustness against impurities
Intriguingly, the traditional picture has recently been challenged by the observation of unconventional superconductors that are unusually robust against non-magnetic doping. For instance, the superconductor Cu$_x$(PbSe)$_5$(Bi$_2$Se$_3$)$_6$ (CPSBS)~\cite{Sasaki:2014}, showing nematic properties and a nodal gap structure~\cite{Andersen:2018}, survives scattering rates much larger than anticipated by Anderson's theorem~\cite{Andersen:2020}. Other materials in the same family,  Cu$_x$Bi$_2$Se$_3$~\cite{Hor:2010, Yonezawa:2017, Kriener:2012} and Nb$_x$Bi$_2$Se$_3$~\cite{Qiu:2015, Smylie:2016, Smylie:2017},  also display unusual robustness against non-magnetic disorder. In-doped SnTe remarkably shows a larger critical temperature for samples with high residual resistivity \cite{Novak:2013}. Irradiated PdTe$_2$ also shows robustness against non-magnetic disorder, with its critical temperature suppressed at a rate that is about sixteen times slower than predicted by standard estimations \cite{Timmons:2020}.

%Theory developed to explain robustness
The first theories developed to address this unusual robustness were
based on the presence of strong spin-orbit coupling (SOC) 
\cite{Michaeli:2012,Nagai:2015}. Later, it became clear that for complex superconductors  (with extra internal degrees of freedom  such as orbitals or sublattices), the concept of superconducting fitness allows for a generalization of Anderson's theorem, providing an universal framework and explanation for the unusual robustness of unconventional superconducting states \cite{Andersen:2020,Timmons:2020}. Specific results for Cu-doped Bi$_2$Se$_3$ report that the robustness of the superconducting state depends not only on the superconducting order parameter, but on details of the electronic structure in the normal state \cite{Sato:2020}. Recently, this understanding was  corroborated by a more complete analysis of the sensitivity of pairing states to various scattering potentials in two-orbital systems \cite{Dentelski:2020, Cavanagh:2021}.

%Microscopic properties of doped-Bi2Se3
Here we focus on superconductors derived from Bi$_2$Se$_3$. Common among these materials is the basic crystallographic unit consisting of quintuple layers (Se - Bi - Se$''$ - Bi$'$ - Se$'$) \cite{Liu:2010},  as schematically depicted in Fig.~\ref{fig:states}. Superconductivity emerges in these systems only after doping, which intrinsically also introduces disorder. The location of impurities within these layers has attracted some interest, since evidence has accumulated that their distribution, and not just the electron donation, plays a decisive role for the formation of unconventional superconductivity~\cite{Lin:position:2021}. Density-functional theory calculations suggest that the most energetically favorable location for  Cu~\cite{WangDFT:2011} and Sr~\cite{LiDFT:2018} dopants should be between the quintuple layers, i.e. in the van der Waals (vdW) gap. X-ray diffraction experiments seem to support this suggestion, since an expansion of the $c$-axis has been observed both in Cu$_x$Bi$_2$Se$_3$~\cite{Shirasawa:2014, Wang:2016, Frohlich:2020} and Sr$_x$Bi$_2$Se$_3$~\cite{Shruti:2015, Liu:2015}. However, dopants in the vdW gap have never been directly detected so far, neither with neutron-scattering experiments in the case of Cu$_x$Bi$_2$Se$_3$~\cite{Frohlich:2020} nor with transmission electron microscopy in the case of Sr$_x$Bi$_2$Se$_3$~\cite{LiDFT:2018}. In the latter compound the vertical position of the impurities has only recently been determined using normal incidence x-ray standing wave measurements~\cite{Lin:position:2021}. It was found that the dopants lie close to the Se and Se$'$ sites with a small vertical displacement towards the center of the quintuple layer, hence not in the vdW gap. 

%Aim of this paper
Motivated by these observations, we investigate in detail how the location of impurities influence the renormalization of the critical temperature in  layered superconductors. Using Bi$_2$Se$_3$ as our example, we derive the scattering matrices for four different scenarios: substitutional (on-site), interstitial (in-between two sites), intercalated (in the vdW gap), and polar defects, as illustrated in Fig. \ref{fig:distr}. We start with an effective microscopic model for the electronic states in the quintuple layers and analyze the effects of impurities in case different pairs of orbitals dominate the effective low-energy electronic structure. Here we are guided by the concept of superconducting fitness and the generalized Anderson's theorem to discuss the robustness of superconducting states in multiple scenarios.

This paper is organized as follows. In Section \ref{sec:micro} we review the microscopic description of the electronic structure of materials in the family of doped Bi$_2$Se$_3$ and the possible s-wave superconducting states. In Section \ref{sec:impurities} after modelling different impurity configurations in the layer basis we discuss how these manifest in effective two-orbital models. In Section \ref{sec:renorm} we then evaluate the effective scattering rates for different impurity distributions and effective model scenarios. Finally, in Section \ref{sec:conclusion} we summarize our findings and how these can help us understand the unusual robustness of superconducting states in layered materials.

%%%%%%%%%%%%%%%%%%%%%%%%%%%%%%%%%%%%%%
\section{Effective microscopic description}\label{sec:micro}
In this section, we introduce and discuss the microscopic structure of the normal-state Hamiltonian and the superconducting order parameters. 

\subsection{Normal state}
\begin{figure}[t!]
	\includegraphics[width=0.6\columnwidth]{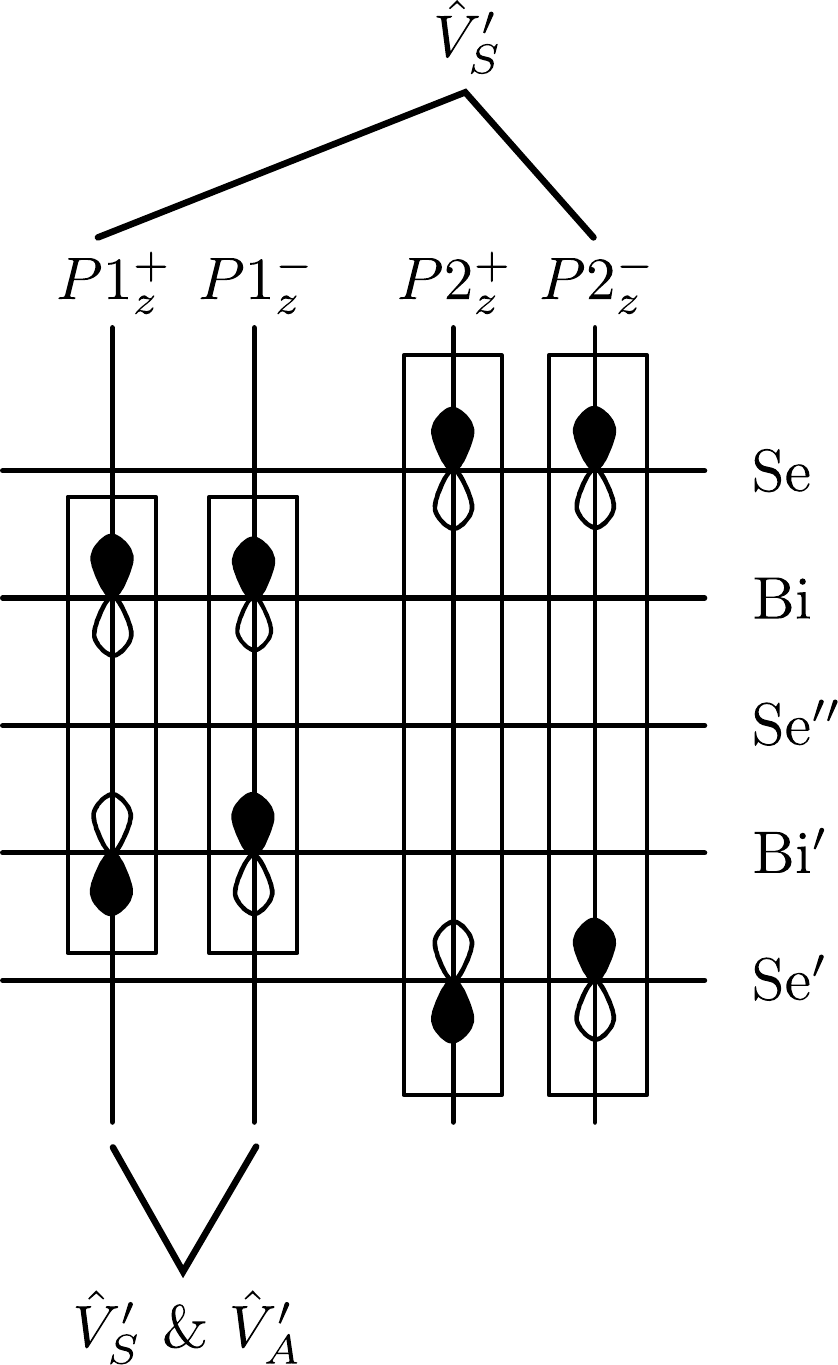}
	\caption{Side view of the quintuple layer unit cell of Bi$_2$Se$_3$. We illustrate the  $PI_{z}^{\alpha}$ orbitals as symmetric ($\alpha=+$) or anti-symmetric ($\alpha=-$) combinations of the $p_z$ orbitals in each layer. We also highlight which scattering matrices remain relevant, if two specific states are chosen as the low-energy basis. As examples, we consider the combinations $\{ P1_{z}^{+}, P2_{z}^{-}\}$ (top) and $\{ P1_{z}^{+}, P1_{z}^{-}\} $ (bottom). More  details on the matrix structures are given in~Table~\ref{tab:tableI}. 
	}
	\label{fig:states}
\end{figure}

Motivated by the phenomenology of Bi$_2$Se$_3$-based superconductors, we start modelling the electronic structure in the quintuple-layer of these materials. First principles calculations suggest that the main orbitals contributing to the low-energy electronic structure of Bi$_2$Se$_3$ stem from $p_z$ orbitals located at the four outermost layers (Se - Bi - Bi$'$ - Se$'$), here labelled as $\ket{S_z}$, $\ket{B_z}$, $\ket{B'_z}$, $\ket{S'_z}$, respectively ~\cite{Liu:2010}. It is convenient to combine these in bonding and anti-bonding configurations,
\begin{align}
    \ket{P1_z^{\pm}} &= \frac{1}{\sqrt{2}} (\ket{B_z} \mp \ket{B_z'}), \label{eqn:P1}\\
    \ket{P2_z^{\pm}} &= \frac{1}{\sqrt{2}} (\ket{S_z} \mp \ket{S_z'}), \label{eqn:P2}
\end{align}
which are schematically indicated in Fig.~\ref{fig:states}. Note that the upper index corresponds to the parity of the state. 

The states written down in Eqs.~(\ref{eqn:P1},~\ref{eqn:P2}) constitute what we call the ``orbital'' basis, denoted by $\{P1_z^{+},~P1_z^{-},~P2_z^{+},~P2_z^{-}\}$. In the ``layer'' basis \{Se,~Bi,~Bi$'$,~Se$'$\} they take the form
\begin{align}
    \ket{P1_z^{\pm}} = \frac{1}{\sqrt{2}} 
    \begin{pmatrix}
    0 \\ 1 \\ \mp 1 \\ 0
    \end{pmatrix},~ 
    \ket{P2_z^{\pm}} = \frac{1}{\sqrt{2}} 
    \begin{pmatrix}
    1 \\ 0 \\ 0\\ \mp 1 
    \end{pmatrix}.
\end{align}
The unitary transformation matrix corresponding to this basis change is
\begin{align}
    \hat{U} = \frac{1}{\sqrt{2}} \begin{pmatrix}
        0 & 0 & 1 & 1 \\
        1 & 1 & 0 & 0  \\
        -1 & 1 & 0 & 0  \\
        0 & 0 & -1 & 1  
    \end{pmatrix}, \label{eqn:unitary}
\end{align}
which will be important in Sect.~\ref{sec:impurities}, where we analyze the structure of different impurity configurations in different reduced orbital bases. 

For Bi$_2$Se$_3$-based materials, it is clear from first principles that only two of these orbitals contribute to the Fermi surface. In particular, it is suggested that these orbitals are $\{P1_z^+, P2_z^-\}$ ~\cite{Liu:2010}. More generally, we can think of any pair of orbitals effectively contributing to the Fermi surface. There are three fundamentally distinct scenarios: i) two orbitals with opposite parity (OP) coming from different types of atoms (this would be the case of Bi$_2$Se$_3$-based materials just mentioned, but also $\{P2_z^+, P1_z^-\}$); ii) two orbitals with OP coming from the same type of atoms ($\{P1_z^+, P1_z^-\}$ or $\{P2_z^+, P2_z^-\}$); iii) two orbitals with equal parity (EP), necessarily associated with different atoms ($\{P1_z^+, P2_z^+\}$ or $\{P1_z^-, P2_z^-\}$). 

For all these cases the low-energy Hamiltonian of the system can be parametrized as
\begin{align}
\hat{H}(\bm{k})= \sum_{a,b}  h_{ab}(\bm{k}) (\hat{\tau}_a\otimes\hat{\sigma}_b), \label{eqn:NS_Ham}
\end{align}
where $\hat{\tau}_{a}$ and $\hat{\sigma}_{b}$, with $\{a,b\} \in \{0,1,2,3\}$, are Pauli matrices encoding the orbital and spin degrees of freedom, respectively. For the OP scenario, the  symmetry-allowed terms in the Hamiltonian are the ones summarized in Table~\ref{tab:H0CPSBS_1}. Analogously, for the EP scenario, we obtain the terms listed in Table~\ref{tab:H0CPSBS_2}.  Naturally, the explicit momentum dependence of the $h_{ab}(\bm{k})$ terms in the normal-state Hamiltonian depends on the point group symmetry of the system (here $D_{3d}$) and the relative parity of the low-energy states. The explicit form of their momentum dependence around the $\Gamma$ point is given in Tables \ref{tab:H0CPSBS_1} and \ref{tab:H0CPSBS_2}, for the OP and EP scenarios, repectively.

\begin{table}[t]
\begin{center}
    \begin{tabular}{ c  c  c  c }
    \hline \hline
    $(a,b)$\rule{0pt}{4mm} &   Irrep &Process &  $h_{ab}(\bk)$ \\[1mm]  \hline
    $(0,0)$\rule{0pt}{4mm} &    \multirow{2}{*}{$A_{1g}$} &\multirow{2}{*}{intra-orbital hopping} & $C_0 + C_1 k_z^2 + C_2 (k_x^2 + k_y^2)$ \\
    $(3,0)$ &    & &  $M_0 + M_1 k_z^2+ M_2 (k_x^2 + k_y^2)$ \\[2mm] 
    $(1,3)$ &   $A_{1u}$ & SOC &  $R_1 k_x (k_x^2-3k_y^2)$  \\[2mm] 
    $(2,0)$ &    $A_{2u}$ & inter-orbital hopping & $B_0 k_z $ \\[2mm] 
    $(1,1)$ &    \multirow{2}{*}{$E_{u}$} & \multirow{2}{*}{SOC}  & $- A_0 k_y $  \\ 
    $(1,2)$ &     & & $ A_0 k_x$ \\[1mm] \hline \hline
        \end{tabular}
        \end{center}
    \caption{  \label{tab:H0CPSBS_1} Parametrization of the normal-state Hamiltonian [Eq.~\eqref{eqn:NS_Ham}] for materials in the family of Bi$_2$Se$_3$ assuming orbitals with OP. For each pair of indexes $(a,b)$ corresponding to the basis matrices $\hat{\tau}_a \otimes \hat{\sigma}_b$, the table highlights the irreducible representations (Irrep) and the physical process that originates them. The last column gives the expansion of the form factors $h_{ab}(\bk)$ for small momentum. %The analogous parametrization obtained for orbitals with EP is given in Table~\ref{tab:H0CPSBS_2}. 
    }
\end{table}
\begin{table}[t]
\begin{center}
    \begin{tabular}{ c  c  c  c }
    \hline \hline
    $(a,b)$\rule{0pt}{4mm} &   Irrep &Process &  $h_{ab}(\bk)$ \\[1mm]  \hline
    $(0,0)$\rule{0pt}{4mm} &    \multirow{3}{*}{$A_{1g}$} &\multirow{2}{*}{intra-orbital hopping} & $C'_0 + C'_1 k_z^2 + C'_2 (k_x^2 + k_y^2)$ \\
    $(3,0)$ &    & &  $M'_0 + M'_1 k_z^2+ M'_2 (k_x^2 + k_y^2)$ \\ 
    $(1,0)$ & & inter-orbital hopping &  $N'_0 + N'_1 k_z^2+ N'_2 (k_x^2 + k_y^2)$  \\[2mm] 
    $(2,3)$ &    $A_{2g}$ & SOC & $R'_1 k_x k_z (k_x^2 - 3k_y^2)$ \\[2mm] 
    $(2,1)$ &    \multirow{2}{*}{$E_{g}$} & \multirow{2}{*}{SOC}  & $- A'_0 k_y k_z$  \\ 
    $(2,2)$ &     & & $ A'_0 k_x k_z$ \\[1mm] \hline \hline
        \end{tabular}
        \end{center}
    \caption{  \label{tab:H0CPSBS_2} Parametrization of the normal-state Hamiltonian [Eq.~\eqref{eqn:NS_Ham}] for the EP scenario. Same description as in Table \ref{tab:H0CPSBS_1}.
    }
\end{table}

\subsection{Superconducting state}
The order parameter of the superconducting state can be written as
\begin{align}
\hat{\Delta}(\bm{k})= \sum_{a,b}  d_{ab}(\bm{k}) \left[\hat{\tau}_a\otimes\hat{\sigma}_b\left(i\hat{\sigma}_2\right)\right]. \label{eqn:gap_func}
\end{align}
From now on we implicitly assume that the order parameter is normalized, i.e. $\braket{||\hat{\Delta}(\bm{k})||^2}_{\rm FS} = 1$, where $\braket{\dots}_{\rm FS}$ denotes the average over the FS, and $||\hat{M}||^2 = \rm{Tr} [\hat{M}\hat{M}^\dagger]/4$ is the Frobenius norm of the matrix $\hat{M}$. Fermionic anti-symmetry requires $\hat{\Delta}(\bk) = -\hat{\Delta}^T(-\bk)$, what means that even-$\bk$ (odd-$\bk$) order parameters are necessarily accompanied by anti-symmetric (symmetric) matrices. For simplicity, in the following we restrict ourselves to $\bm{k}$-independent $d_{ab}(\bm{k})$, also referred to as s-wave superconducting states. This reduces the space of order parameters to the six anti-symmetric basis matrices listed in Table~\ref{tab:SC}. Note that they are labelled as $[a,b]$ (rectangular brackets are used to label the superconducting states, while round brackets label the terms in the normal-state Hamiltonian) and that they are independent of the choice of EP or OP orbitals contributing to the Fermi surface. However, as summarized in Table~\ref{tab:SC}, the parity of the orbitals influences the symmetry of the order parameter and its associated irreducible representation.
\begin{table}[t!]
\begin{center}
    \begin{tabular}{ c  c  c  c  c  }
    \hline \hline
    $[a,b]$\rule{0pt}{4mm} & Spin &  Orbital &  Irrep~(OP)   & Irrep~(EP)  \\[1mm] \hline
    $[0,0]$\rule{0pt}{4mm}& 
     \multirow{2}{*}{Singlet}&
          \multirow{2}{*}{Trivial}&
                \multirow{2}{*}{$A_{1g}$}&  \multirow{2}{*}{$A_{1g}$} \\ 
     	$[3,0]$ & & & &   \\[2mm]
		$[2,3]$  & Triplet  & Singlet & {$A_{1u}$} & {$A_{2g}$}   \\[2mm]
		$[1,0]$  & Singlet  & Triplet & {$A_{2u}$} & {$A_{1g}$}   \\[2mm]
		$[2,1]$ &
     \multirow{2}{*}{Triplet}&
          \multirow{2}{*}{Singlet}&
              \multirow{2}{*}{$E_{u}$} &  \multirow{2}{*}{$E_{g}$}
				\\ 
	    $[2,2]$ &&& &  \\[1mm] \hline \hline
    \end{tabular}
               \end{center}
        \caption{\label{tab:SC} Momentum-independent superconducting order parameters for two-orbital models. We highlight here the spin and orbital characters as well as the irreducible representation of the respective gap matrix for the odd parity (OP) and even parity (EP) cases.}
\end{table}

Having clarified our microscopic model, we consider in the following different scattering potentials out of non-magnetic impurities.

%%%%%%%%%%%%%%%%%%%%%%%%%%%%%%%%%%%%%%
\section{Impurity configurations in the quintuple layer and orbital bases}\label{sec:impurities}

\begin{figure}[t]
	\includegraphics[width=0.8\columnwidth]{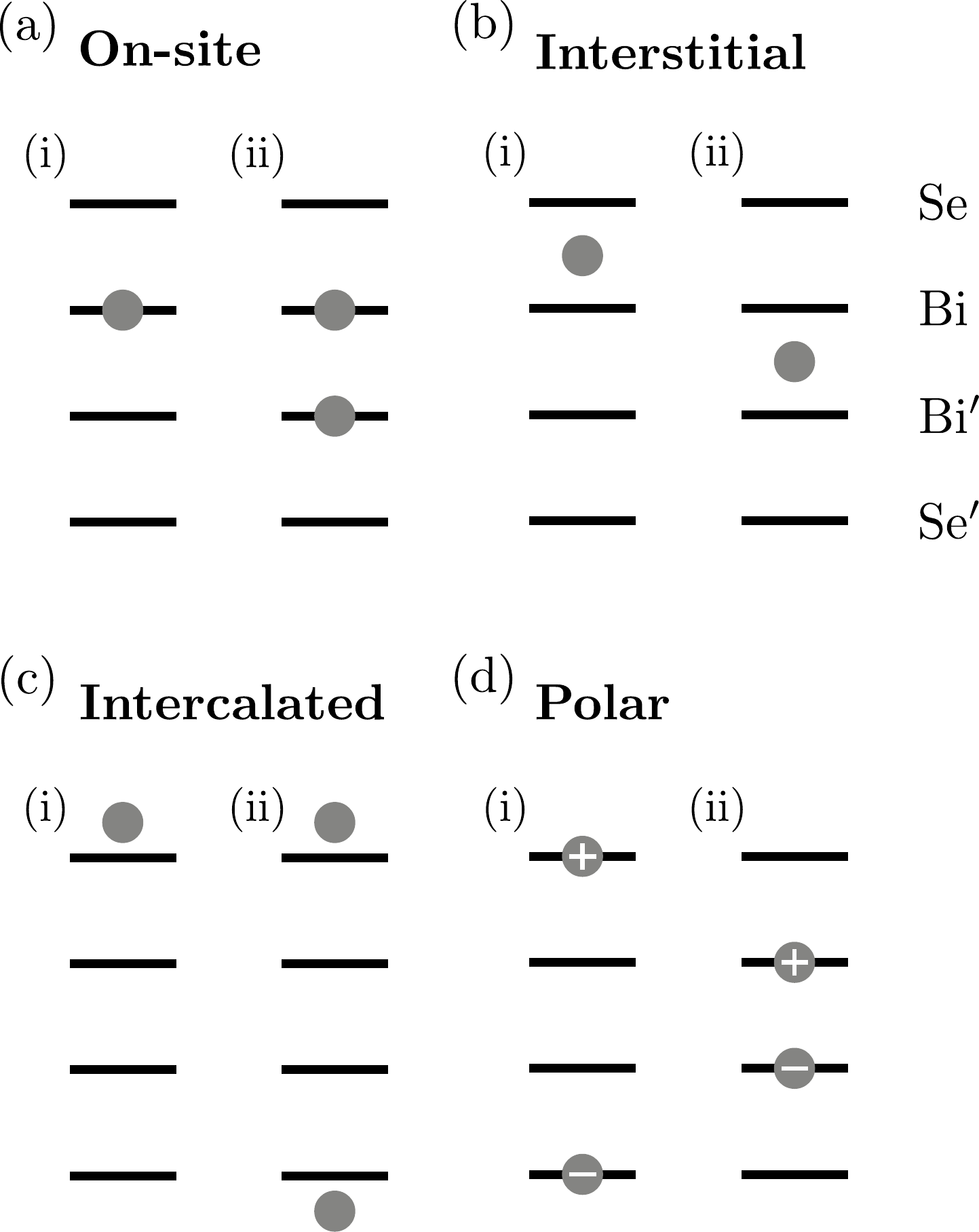}
	\caption{Schematic illustration of (a) on-site, (b) interstitial, (c) intercalated and (d) polar impurity configurations (marked as dark grey circles). 
	In Eqs.~(\ref{eqn:decomp_on}) to (\ref{eqn:decomp_ic}) we show that the scattering potential can be always decomposed into a superposition of a symmetric and an anti-symmetric impurity potentials. The polar configuration consists of oppositely charged impurities in opposite layers, which leads to a fully anti-symmetric potential [Eq.\eqref{eqn:decomp_po}]. 
	}
	\label{fig:distr}
\end{figure}

To unveil the effects of the impurity location in materials belonging to the family of Bi$_2$Se$_3$, we distinguish between on-site (on), interstitial (is), intercalated (ic) and polar (po) configurations, which are illustrated in Fig.~\ref{fig:distr}.

The scattering matrices written in the layer basis \{Se,~Bi,~Bi$'$,~Se$'$\} have an entry on the diagonal, if the corresponding layer is affected by the impurity. For instance, the on-site configurations shown in Fig.~\ref{fig:distr}~(a) are given by
\begin{align}
    \hat{V}_{\rm on}^{\rm (i)} = 
    \begin{pmatrix}
        0 & 0 & 0 & 0 \\
        0 & V & 0 & 0 \\
        0 & 0 & 0 & 0 \\
        0 & 0 & 0 & 0 
    \end{pmatrix},~ 
    \hat{V}_{\rm on}^{\rm (ii)} = 
\begin{pmatrix}
        0 & 0 & 0 & 0 \\
        0 & V & 0 & 0 \\
        0 & 0 & V & 0 \\
        0 & 0 & 0 & 0 
    \end{pmatrix},
    \label{eqn:oN_{uc}ite}
\end{align}
where $V > 0$ denotes the impurity scattering strength, which we take to be the same for all impurities. If the impurities are located between the Bi and Se sites, as in Fig.~\ref{fig:distr}~(b), we assume that their presence affects both neighboring layers, which leads to
\begin{align}
    \hat{V}_{\rm is}^{\rm (i)} = \begin{pmatrix}
        V & 0 & 0 & 0 \\
        0 & V & 0 & 0  \\
        0 & 0 & 0 & 0  \\
        0 & 0 & 0 & 0  
    \end{pmatrix}, ~
    \hat{V}_{\rm is}^{\rm (ii)} = \begin{pmatrix}
        0 & 0 & 0 & 0 \\
        0 & V & 0 & 0  \\
        0 & 0 & V & 0  \\
        0 & 0 & 0 & 0  
    \end{pmatrix}. 
    \label{eqn:intersti}
\end{align}
Correspondingly, we find for the intercalated scenario, depicted in Fig.~\ref{fig:distr}~(c),
\begin{align}
    \hat{V}_{\rm ic}^{\rm (i)} = \begin{pmatrix}
        V & 0 & 0 & 0 \\
        0 & 0 & 0 & 0  \\
        0 & 0 & 0 & 0  \\
        0 & 0 & 0 & 0  
    \end{pmatrix}, ~
    \hat{V}_{\rm ic}^{\rm (ii)} = \begin{pmatrix}
        V & 0 & 0 & 0 \\
        0 & 0 & 0 & 0  \\
        0 & 0 & 0 & 0  \\
        0 & 0 & 0 & V  
    \end{pmatrix},
    \label{eqn:intercal}
\end{align}
and for the polar case, exemplified in Fig.~\ref{fig:distr}~(d),
\begin{align}
    \hat{V}_{\rm po}^{\rm (i)} = \begin{pmatrix}
        V & 0 & 0 & 0 \\
        0 & 0 & 0 & 0  \\
        0 & 0 & 0 & 0  \\
        0 & 0 & 0 & -V  
    \end{pmatrix}, ~
    \hat{V}_{\rm po}^{\rm (ii)} = \begin{pmatrix}
        0 & 0 & 0 & 0 \\
        0 & V & 0 & 0  \\
        0 & 0 & -V & 0  \\
        0 & 0 & 0 & 0  
    \end{pmatrix}.
    \label{eqn:polar}
\end{align}

There are two linearly independent symmetric (S) configurations, which are even under inversion. They take the form 
\begin{gather}
    \hat{V}_{\rm S, 1} = \begin{pmatrix}
        V & 0 & 0 & 0 \\
        0 & 0 & 0 & 0 \\
        0 & 0 & 0 & 0 \\
        0 & 0 & 0 & V 
    \end{pmatrix},~\hat{V}_{\rm S, 2} = \begin{pmatrix}
        0 & 0 & 0 & 0 \\
        0 & V & 0 & 0 \\
        0 & 0 & V & 0 \\
        0 & 0 & 0 & 0 
    \end{pmatrix}. \label{eqn:sym_matr}
\end{gather}
The anti-symmetric (A) counterparts are given by 
\begin{gather}
    \hat{V}_{\rm A, 1} = \begin{pmatrix}
        V & 0 & 0 & 0 \\
        0 & 0 & 0 & 0 \\
        0 & 0 & 0 & 0 \\
        0 & 0 & 0 & -V 
    \end{pmatrix},~\hat{V}_{\rm A, 2} = \begin{pmatrix}
        0 & 0 & 0 & 0 \\
        0 & V & 0 & 0 \\
        0 & 0 & -V & 0 \\
        0 & 0 & 0 & 0 
    \end{pmatrix}. \label{eqn:asym_matr}
\end{gather}
Every impurity configuration can be split up into symmetric and anti-symmetric parts using Eqs.~(\ref{eqn:sym_matr}, \ref{eqn:asym_matr}). For example, the configurations introduced above can be rewritten as
\begin{gather}
    \begin{gathered}
    \hat{V}_{\rm on}^{\rm (i)}
    = \frac{1}{2} \left(\hat{V}_{\rm S, 2} + \hat{V}_{\rm A, 2}\right),  \\
    \hat{V}_{\rm on}^{\rm (ii)}
    = \hat{V}_{\rm S, 2} , \label{eqn:decomp_on} 
    \end{gathered}\\[2mm]
    \begin{gathered}
    \hat{V}_{\rm is}^{\rm (i)}
    = \frac{1}{2} \left( \hat{V}_{\rm S, 1} + \hat{V}_{\rm S, 2} 
    + \hat{V}_{\rm A, 1} + \hat{V}_{\rm A, 2} \right),\\
    \hat{V}_{\rm is}^{\rm (ii)}
    = \hat{V}_{\rm S, 2} , \label{eqn:decomp_is} 
    \end{gathered}\\[2mm]
    \begin{gathered}
    \hat{V}_{\rm ic}^{\rm (i)}
    = \frac{1}{2} \left(\hat{V}_{\rm S, 1} + 
     \hat{V}_{\rm A, 1}\right),  \\
    \hat{V}_{\rm ic}^{\rm (ii)}
    = \hat{V}_{\rm S, 1} , \label{eqn:decomp_ic}
    \end{gathered}
\end{gather}
and 
\begin{gather}
    \begin{gathered}
    \hat{V}_{\rm po}^{\rm (i)}
    = \hat{V}_{\rm A, 1}, ~\hat{V}_{\rm po}^{\rm (ii)}
    = \hat{V}_{\rm A, 2}. \label{eqn:decomp_po}
    \end{gathered}
\end{gather}

\begin{figure}[t!]
	\includegraphics[width=0.68\columnwidth]{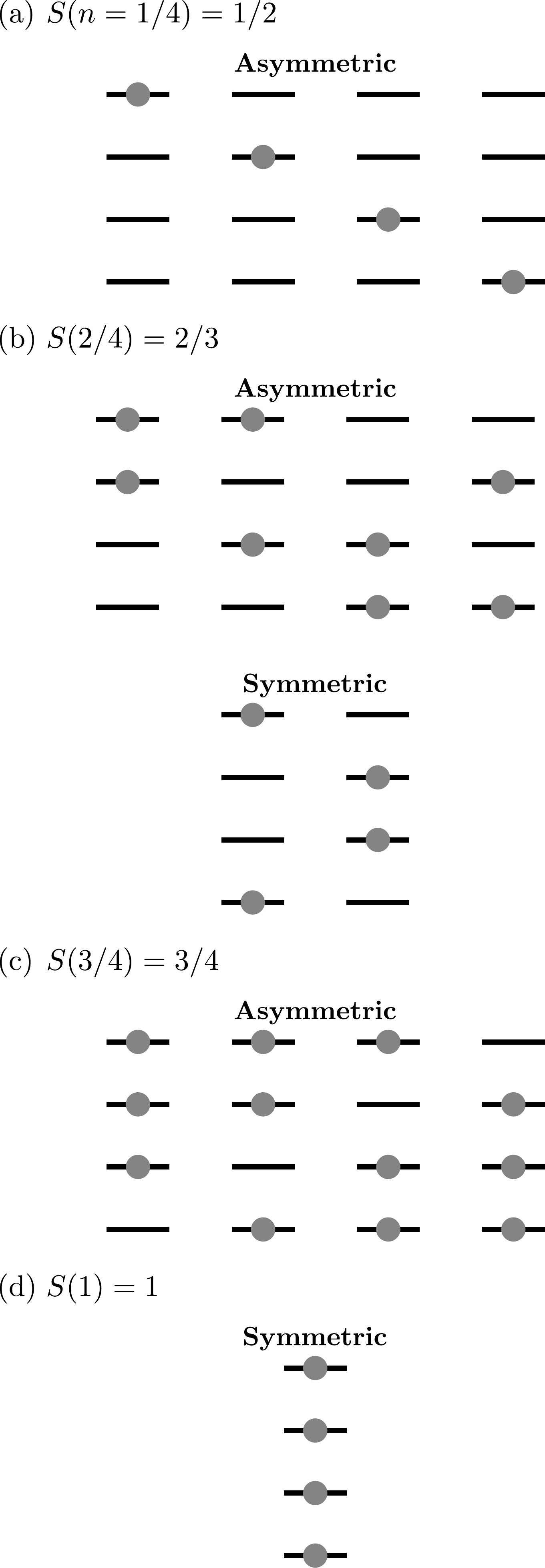}
	\caption{Overview of all possible on-site impurity configurations for a single unit cell with four layers. The maximum number of impurities (indicated by the grey circles) is $N_{\rm max}=4$. For each number of impurities $N$ we included the corresponding symmetric share~$S(n=N/N_{\rm max})$.
	}
	\label{fig:share_illustr}
\end{figure}

\begin{figure}[t!]
	\includegraphics[width=0.99\columnwidth]{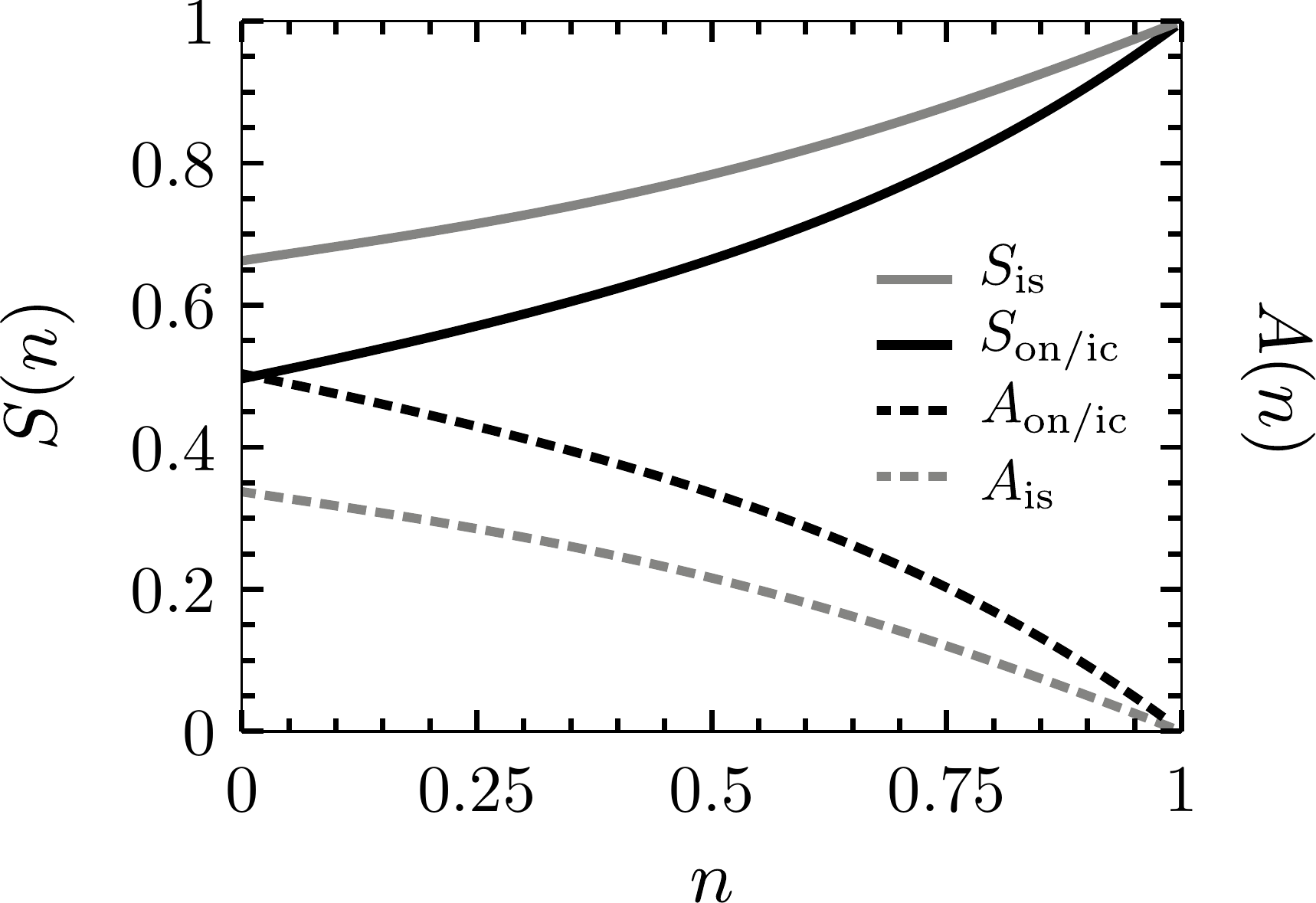}
	\caption{Evolution of the symmetric share of impurity configurations, $S(n)$ (solid lines), as a function of the filling $n=N/N_{\rm max}$ for $N_{uc} \gg 1$. The difference between on-site (on),  interstitial (is), and intercalated (ic)  configurations is illustrated in Fig.~\ref{fig:distr}. The dashed lines represent the anti-symmetric shares given by $A(n)=1-S(n)$.}
	\label{fig:shares_distr}
\end{figure}

Using this framework, we can decompose every possible impurity distribution for a single unit cell in terms of symmetric and anti-symmetric scattering matrices. Assuming a system with $N_{uc}$ unit cells, each with four layers, the maximum number of impurities is equal to $N_{\rm max} = 4N_{uc}$, $3N_{uc}$ or $2N_{uc}$ for on-site, interstitial or intercalated and polar configurations, respectively. For a fixed number of impurities it is a straightforward combinatorial problem to find all possible impurity configurations. To give a concrete example, we show all possible, on-site configurations for $N_{uc} = 1$ in Fig.~\ref{fig:share_illustr}.
For a single impurity, $n=N/N_{\rm max}=1/4$, there are only asymmetric configurations. These can be decomposed into a sum of symmetric and anti-symmetric potentials, what leads to a symmetric share $S(n)=1/2$. The symmetric share increases with growing $N$ until it reaches one for the completely filled unit cell ($n=1$). By calculating the symmetric share numerically for all different configurations and for an arbitrary large number of sites we find that $S(n)$ quickly converges to a universal result, which is depicted by the solid lines in Fig.~\ref{fig:shares_distr}. The dashed lines represent the anti-symmetric shares $A(n)=1-S(n)$. Note that for polar impurities, whose scattering potential is completely anti-symmetric, we simply obtain $A_{\rm po}(n)=1$.
Furthermore, as indicated in Fig.~\ref{fig:shares_distr}, the shares for on-site and intercalated impurities evolve in the same way as a function of the filling $n$.
\par 

We now examine which orbital wave functions are affected by the different impurity scattering matrices. To this extent we transform the symmetric and anti-symmetric matrices [Eqs.~(\ref{eqn:sym_matr},~\ref{eqn:asym_matr})] to the $\{P1_z^{+},~P1_z^{-},~P2_z^{+},~P2_z^{-}\}$ basis by evaluating $\hat{U}^{\dagger}\hat{V}\hat{U}$, with $\hat{U}$ given by Eq.~\eqref{eqn:unitary}. We obtain for the symmetric scattering matrices, 
\begin{gather}\label{eqn:orbital_S}
    \hat{V}'_{\rm S, 1} = \begin{pmatrix}   
        0 & 0 & 0 & 0 \\
        0 & 0 & 0 & 0 \\
        0 & 0 & V & 0 \\
        0 & 0 & 0 & V 
    \end{pmatrix},~\hat{V}'_{\rm S, 2} = \begin{pmatrix}
        V & 0 & 0 & 0 \\
        0 & V & 0 & 0 \\
        0 & 0 & 0 & 0 \\
        0 & 0 & 0 & 0 
    \end{pmatrix}, 
\end{gather}
while the purely anti-symmetric matrices are 
\begin{gather}\label{eqn:orbital_A}
    \hat{V}'_{\rm A, 1} = \begin{pmatrix}
        0 & 0 & 0 & 0 \\
        0 & 0 & 0 & 0 \\
        0 & 0 & 0 & V \\
        0 & 0 & V & 0 
    \end{pmatrix},~\hat{V}'_{\rm A, 2} = \begin{pmatrix}
        0 & V & 0 & 0 \\
        V & 0 & 0 & 0 \\
        0 & 0 & 0 & 0 \\
        0 & 0 & 0 & 0 
    \end{pmatrix}.
\end{gather}
As discussed in the previous section, the ultimate low-energy description of the electronic states can be cast in terms of two orbitals. From Eqs.~(\ref{eqn:orbital_S}) and (\ref{eqn:orbital_A}), we see that the effective scattering matrices are going to depend on the choice of orbitals. In particular, some of the scattering matrices vanish in the low-energy basis. A summary of the reduced scattering matrices for all possible choices of two low-energy states is given in Table~\ref{tab:tableI}.
\begin{table}[t!]
	\centering
	\begin{tabular}{c  c  c  c  c}
	 \hline
     \hline
     Two-orbital basis\rule{0pt}{4mm} & $\hat{V}'_{\rm S, 1}$ & $\hat{V}'_{\rm S, 2}$ & $\hat{V}'_{\rm A, 1}$ & $\hat{V}'_{\rm A, 2}$ \\[1mm]
     \hline
     $\{ P1_{z}^{+}, P1_{z}^{-}\} $\rule{0pt}{4mm}& $0$ & $ \hat{\tau}_0$ & 0 & $ \hat{\tau}_1$  \\[1.6mm]
     $\{ P2_{z}^{+}, P2_{z}^{-}\} $ & $\hat{\tau}_0$ & 0 & $ \hat{\tau}_1$ & 0  \\[1mm]
     $\{ P1_{z}^{\alpha}, P2_{z}^{\alpha'}\} $\rule{0pt}{3.5mm}& $\hat{\tau}_0-\hat{\tau}_3$ & $ \hat{\tau}_0+\hat{\tau}_3$ & 0 & 0  \\[1.5mm]
	 \hline
     \hline
	\end{tabular}
	  \caption{Explicit form of the symmetric and anti-symmetric scattering matrices for a particular two-orbital basis chosen from $\{P1_z^{+},~P1_z^{-},~P2_z^{+},~P2_z^{-}\}$, with $\alpha,\alpha' \in \{+,-\}$. Pauli matrices acting in orbital space are denoted by $\hat{\tau}_{i}$ with $i \in \{0,1,2,3\}$, where $\hat{\tau}_0$ corresponds to the identity matrix.
	  }
		\label{tab:tableI}
\end{table}
Note that for orbitals involving different types of atoms, $\{ P1_{z}^{\alpha}, P2_{z}^{\alpha'}\}$, all anti-symmetric scattering matrices are zero in the low-energy basis. On the contrary, if the combinations involve the same type of atom, meaning $\{ P1_{z}^{+}, P1_{z}^{-}\}$ and $\{ P2_{z}^{+}, P2_{z}^{-}\}$, both the symmetric and anti-symmetric matrices are finite, as illustrated in Fig.~\ref{fig:states}. 
In the next section, we will analyze the implications of these findings regarding the renormalization of the superconducting temperature.

%%%%%%%%%%%%%%%%%%%%%%%%%%%%%%%
\section{Renormalization of $\boldsymbol{T_{c}}$} \label{sec:renorm}

Within the standard self-consistent Born approximation, the reduction of the critical temperature in the presence of impurities for simple superconductors can be cast as~\cite{mineev1999introduction}:
\begin{align}
    \log \left(\frac{T_c}{T_{c0}}\right) = \Psi\left( \frac{1}{2} \right) - \Psi\left( \frac{1}{2} + \frac{1}{4\pi \tau_{\rm eff} T_c} \right), \label{eqn:Tc_renorm}
\end{align}
with $\Psi(x)$ denoting the digamma function and $\tau_{\rm eff}$ the effective scattering rate defined as
\begin{align}
  \tau_{\rm eff}^{-1} = \tau_{\rm n}^{-1} - \tau_{\rm sc}^{-1}, 
\end{align}
in terms of the normal ($\tau_{\rm n}$) and superconducting ($\tau_{\rm sc}$) scattering rates. Below we begin by discussing how $\tau_{\rm n}$ and $\tau_{\rm sc}$ are obtained for complex superconductors with two orbital degrees of freedom contributing to a single Fermi surface. In Subsect.~\ref{sec:case_study}, we then discuss different scenarios with specific choices of orbitals and impurity location.

\subsection{The normal scattering rate}
The influence of impurity scattering on the electronic system is described by the renormalization of the Green's function of the system, which we calculate by solving Dyson's equation~\cite{maki2018gapless},
\begin{align}
    \hat{G}^{-1}(\bm{k}, i\omega_n) = \hat{G}_0^{-1}(\bm{k}, i\omega_n) - \hat{\Sigma}_1(\bm{k}, i\omega_n).
\end{align}
Here we introduced the bare Green's function $\hat{G}_0(\bm{k}, i\omega_n)$ and self-energy $\hat{\Sigma}_1(\bm{k}, i\omega_n)$ of the normal state, which depend in general on the wave vector $\bm{k}$ and the fermionic Matsubara frequencies $\omega_n = (2n+1)\pi k_B T$. The bare Green's function is defined as  
\begin{align}
    \hat{G}^{-1}_0(\bm{k}, i\omega_n) = i\omega_n (\hat{\tau}_0 \otimes \hat{\sigma}_0)- \hat{H}(\bm{k}), \label{eqn:bare_GF}
\end{align}
with $\hat{H}(\bm{k})$ given by Eq.~\eqref{eqn:NS_Ham}. Using the Born approximation and assuming isotropic scattering, the self-energy has no $\bm{k}$-dependence and takes the form
\begin{align}
    \hat{\Sigma}_1(i\omega_n) = \sum_i n_i \hat{V}_i \int \frac{d^3 k}{(2\pi)^3} \hat{G}(\bm{k}, i\omega_n) \hat{V}_i,
\end{align}
where $\hat{V}_i$ are the different scattering matrices in the two-orbital basis, as summarized in Table \ref{tab:tableI}.
After inverting Eq.~\eqref{eqn:bare_GF}, inserting the result in the equation for the self-energy and solving the Dyson's equation self-consistently, we end up with a renormalization of the Matsubara frequencies. This allows us to identify the normal-state scattering rate~\cite{Cavanagh:2021},
\begin{align}
    \tau^{-1}_{\rm n} &= \sum_i\frac{\pi}{2} V_i^2 N(0) n_i X_i(n) \nonumber \\
    &\hspace{1cm}\times \Bigg[1+\sum_{a,b} C_{\text{n},i}^{ab}\langle \hat{h}_{ab}(\bm{k})\rangle^2\Bigg], \label{eqn:n_rate}
\end{align}
with $N(0)$ denoting the density of states at the Fermi level, $V_i$ the magnitude of the scattering potential, and $X_i(n)=\{S_i(n), A_i(n)\}$ depending on the specific impurity potential $\hat{V}_i$, as indicated in Table \ref{tab:tableI}. %Note that we drop from now on the FS index of the angular brackets to shorten the notation. 
The factors $C_{\text{n},i}^{ab}$ are equal to $+1$ ($-1$), if the corresponding term $(a,b)$ in the normal-state Hamiltonian  [Eq.~\eqref{eqn:NS_Ham}] commutes (anti-commutes) with the scattering potential $\hat{V}_i$. Here $\langle f(\bm{k})\rangle$ denotes the average over the Fermi surface of the function $f(\bm{k})$. Note that the sum of all symmetry-allowed pairs $(a,b)$ excludes $(0,0)$ and that the coefficients are normalized,
\begin{align}
  \sum_{a,b} \hat{h}^2_{ab}(\bm{k})=1.  
\end{align}
Moreover, we implicitly assume a weak momentum-dependence of the scattering rate such that we can simply consider its FS average. In Table~\ref{tab:Cn} we list the values of $C_{\text{n},i}^{ab}$ for the all scattering potentials $\hat{V}_i = \hat{\tau}_0, \hat{\tau}_1, \hat{\tau}_3$. 
Note that we have to distinguish between OP and EP basis states, since the structure of the normal-state Hamiltonian is different for the EP or OP scenarios, as can be seen from Tables~\ref{tab:H0CPSBS_1} and \ref{tab:H0CPSBS_2}.

\subsection{The superconducting scattering rate}
In the superconducting state, we additionally have to take into account the renormalization of the anomalous Green's function $\hat{F}(\bm{k}, i\omega_n)$, which incorporates the pairing potential. Since we are only interested in the change of $T_c$, we can restrict ourselves to linear order in the gap function. The anomalous Green's function is then simply given by 
\begin{align}
    \hat{F}(\bm{k}, i\omega_n) \approx - \hat{G}(\bm{k}, i\omega_n) &\left[ \hat{\Delta}(\bm{k}) + \hat{\Sigma}_2(i\omega_n) \right]\nonumber \\
    &\hspace{1cm}\times \hat{G}^T(-\bm{k}, -i\omega_n),
\end{align}
where $\hat{\Delta}(\bm{k})$ is given by Eq.~\eqref{eqn:gap_func}. We introduced the self-energy of the superconducting state $\hat{\Sigma}_2(i\omega_n)$, which in the Born approximation is defined as 
\begin{align}
    \hat{\Sigma}_2(i\omega_n) = -\sum_i n_i \hat{V}_i \int \frac{d^3 k}{(2\pi)^3} \hat{F}(\bm{k}, i\omega_n) \hat{V}_i^{\dagger} .
\end{align}
Analogously to before, we calculate first the anomalous Green's function and the corresponding self-energy. Solving then everything self-consistently enables us to find the renormalization of the pairing potential. From here on, we restrict ourselves to purely unconventional states, i.e. the non-$A_{1g}$ gap functions. This allows us to directly obtain a closed-form solutions for the scattering rates. Of course, the framework can also be extended to the $A_{1g}$ channel. However, one then has to account for a superposition of multiple superconducting components, which we want to avoid here. Neglecting interband pairing contributions we finally obtain for the superconducting scattering rate~\cite{Cavanagh:2021},
\begin{align}
    \tau^{-1}_{\rm sc} = \sum_i\frac{\pi}{2} V_i^2 N(0) n_i X_i(n) C_{\text{sc}, i}^{ab}\left[1- \langle F_C \rangle\right], \label{eqn:sc_rate}
\end{align}
where $C_{\text{sc}, i}^{ab}$ is equal to $+1$ ($-1$), if the scattering potential $\hat{V}_i$ commutes (anti-commutes) with the particular gap function labelled as $[a,b]$~[Eq.~\eqref{eqn:gap_func}]. The values for all different $s$-wave superconducting states and scattering potentials are summarized in Table~\ref{tab:Cn}. Note that the EP scenario has less active scattering matrices if compared to the OP scenario. In Eq.~\eqref{eqn:sc_rate} we introduced the normalized average of the Fitness function, 
\begin{align}
   \braket{{F}_C} = \left\langle\frac{||\hat{F}_C(\bm{k})||^2}{\sum_{(a,b) \neq (0,0)} h_{ab}^2(\bm{k})}\right\rangle,
\end{align}
with the Fitness matrix \cite{Ramires:2016,Ramires:2018} given by
\begin{align}
    \hat{F}_C (\bm{k}) &= \hat{H}(\bm{k}) \hat{\Delta}(\bm{k}) - \hat{\Delta}(\bm{k}) \hat{H}^T(-\bm{k}). \label{eqn:fitness}
\end{align}
In Table~\ref{tab:Fc} we provide an overview of the terms $(a,b)$ in the normal-state Hamiltonian, which contribute to a finite Fitness function for the different pairing channels of interest labelled as $[a,b]$. As before, we also include the possibility of OP and EP basis states. 
\begin{table}[t!]
	\centering
	\begin{tabular}{c c c  c }
     \multicolumn{4}{c}{$C_{\text{n},i}^{ab}$\rule{0pt}{5mm}}\\[1mm]
     \hline\hline
     \multirow{2}{*}{$(a,b)$} & \multicolumn{3}{c}{$\hat{V}'_i$\rule{0pt}{4mm}} \\ 
     & $\hat{\tau}_0$ & $\hat{\tau}_1$ & $\hat{\tau}_3$ \\[1mm]
     \hline
     $(3,0)$\rule{0pt}{3.7mm} & $+1$ & $-1$  & $+1$  \\[1mm]
     $(1,0)$ & $+1$ & $+1$ & $-1$  \\[1mm]
     \hline
     \hline
	\end{tabular}
	\hspace{0.5cm}
	\begin{tabular}{c  c  c  c}
     \multicolumn{4}{c}{$C_{\text{sc},i}^{ab}$}\\[1mm]
     \hline
     \hline
     \multirow{2}{*}{$[a,b]$} & \multicolumn{3}{c}{$\hat{V}'_i$\rule{0pt}{4mm}} \\ 
     & $\hat{\tau}_0$ & $\hat{\tau}_1$ & $\hat{\tau}_3$ \\[1mm]
     \hline
     $[0,0]$\rule{0pt}{3.7mm}  & $+1$ & $+1$ & $+1$  \\[1mm]
     $[3,0]$&$+1$ & $-1$ & $+1$  \\[1mm]
     $[2,3]$& $+1$ & $-1$ & $-1$  \\[1mm]
     $[1,0]$& $+1$ & $+1$ & $-1$  \\[1mm]
     $[2,1]$& $+1$ & $-1$ & $-1$  \\[1mm]
     $[2,2]$& $+1$ & $-1$ & $-1$  \\[1mm]
	 \hline\hline
	\end{tabular}
	  \caption{Explicit values of the factors $C_{\text{n},i}^{ab} = \pm 1$ (left) and $C_{\text{sc},i}^{ab} = \pm 1$ (right), which encode the commutation relations between the scattering potentials $\hat{V}'_i$ and  $h_{ab}(\bm{k})$ or $d_{ab}$, respectively. These factors appear in the equations of the normal-state [Eq.~\eqref{eqn:n_rate}] and superconducting [Eq.~\eqref{eqn:sc_rate}] scattering rates. We assume an effective two-orbital model with OP or EP basis states. The scattering matrices in the low-energy model for each of these cases are contained in~Table~\ref{tab:tableI}. Note that in the left table we have only included the $(a,b)$ terms, which can belong to the $A_{1g}$ representation, since they do not vanish after a FS average and ultimately contribute to the normal-state scattering rate. 
	  }
		\label{tab:Cn}
\end{table}
\begin{table}[t!]
	\centering
	\begin{tabular}{c c c}
     \multicolumn{3}{c}{$\hat{F}_{C}^{\rm OP}(\bm{k})$}\\[1mm]
     \hline
     \hline
     $[a,b]$ & Irrep & $(a,b)$\rule{0pt}{4mm} \\[1mm]
     \hline
     $[0,0]$\rule{0pt}{3.7mm}& \multirow{2}{*}{$A_{1g}$} & $-$ \\
     $[3,0]$& & $(2,0),(1,1),(1,2),(1,3)$ \\[1mm]
     $[2,3]$& $A_{1u}$ & $(3,0),(1,3)$ \\[1mm]
     $[1,0]$& $A_{2u}$ & $(3,0),(2,0)$ \\[1mm]
     $[2,1]$& \multirow{2}{*}{$E_{u}$} & $(3,0),(1,2)$ \\
     $[2,2]$& & $(3,0),(1,1)$\\[1mm]
	 \hline\hline
     \multicolumn{3}{c}{$\hat{F}_{C}^{\rm EP}(\bm{k})$\rule{0pt}{5mm}}\\[1mm]
     \hline\hline
     $[0,0]$\rule{0pt}{3.7mm}& \multirow{3}{*}{$A_{1g}$} & $-$ \\
     $[3,0]$& & $(1,0),(2,1),(2,2),(2,3)$ \\
     $[1,0]$& & $(3,0),(2,1),(2,2),(2,3)$ \\[1mm]
     $[2,3]$& $A_{2g}$ & $(1,0),(3,0),(2,1),(2,2)$ \\[1mm]
     $[2,1]$& \multirow{2}{*}{$E_{g}$} & $(1,0),(3,0),(2,2),(2,3)$ \\ 
     $[2,2]$& & $(1,0),(3,0),(2,1),(2,3)$\\[1mm]
	 \hline\hline
	\end{tabular}
	  \caption{The terms $(a,b)$ of the normal-state Hamiltonian [Eq.~\eqref{eqn:NS_Ham}], contributing to the finite Fitness function as $F_C(\bm{k}) = \sum_{a,b} \hat{h}_{ab}^2(\bm{k})$, for the corresponding s-wave superconducting states labelled as $[a,b]$ [Eq.~\eqref{eqn:gap_func}] in both  OP and EP scenarios.
	  }
		\label{tab:Fc}
\end{table}

\subsection{Analysis of three scenarios} \label{sec:case_study}
In the following, we will use these results to analyze qualitatively the impurity-induced renormalization of the superconducting critical temperature for three generic scenarios of multi-layered superconductors. The first example, which is about OP basis states originating from distinct types of atoms, is most closely related to the Bi$_2$Se$_3$-related superconductors. In the second and third examples, we then discuss how the renormalization changes if the states have EP or have OP but originate from the same type of atoms. For all of these instances we consider on-site, interstitial, intercalated and polar impurity configurations.

% case 1 %%%%%%%%%%%%%%%%%%%%%%%%%%%%%%%%%%%%%%%%%%%%%%%%
\subsubsection{OP orbitals from different atoms} 
For concreteness, here we assume the low-energy orbitals are $\{P1_z^+, P2_z^-\}$,  the first associated with an even parity orbital originating from Bi atoms,  the second associated with an odd parity orbital stemming from the Se atoms. From Table~\ref{tab:tableI}, we conclude that only the symmetric impurity configurations contribute to scattering. Thus, our first conclusion is that in the presence of purely polar impurities, which are entirely anti-symmetric, the superconducting state is left untouched.

\begin{figure}[t!]
	\includegraphics[width=0.65\columnwidth]{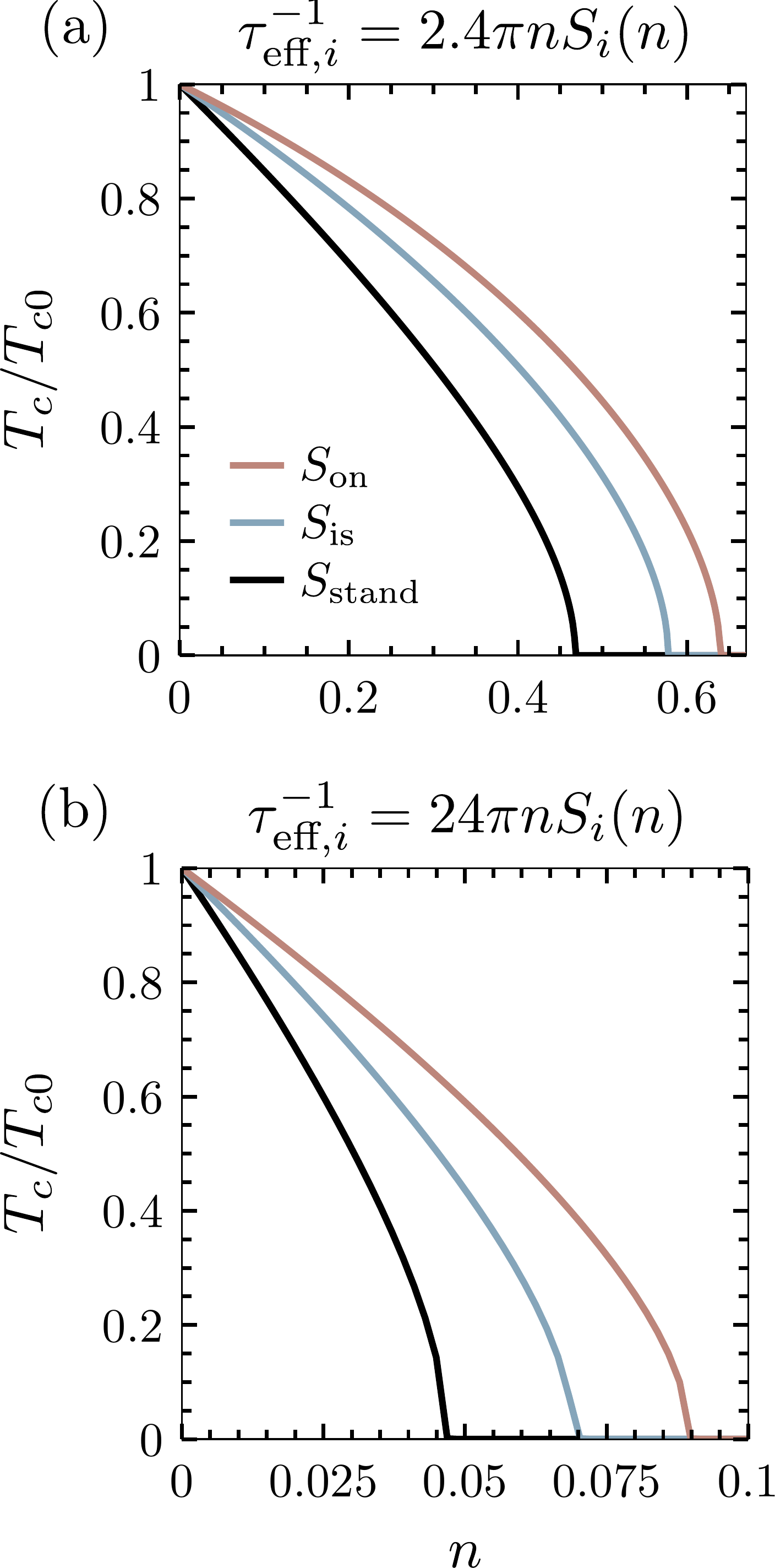}
	\caption{Suppression of the normalized superconducting transition temperature $T_c/T_{c0}$ versus the impurity concentration $n$, if only the symmetric share $S_i(n)$ of impurity distributions contributes to scattering, calculated from Eqs.~(\ref{eqn:Tc_renorm},~\ref{eqn:eff_scat_on&is}). The standard case corresponds to the situation where the share of symmetric configurations stays constant as a function of the filling, i.e. $S_{\rm stand}=1$. For the top (bottom) figure we choose a small (large) prefactor to the scattering rate, corresponding to a weak (strong) scattering potential.
	}
	\label{fig:Tc_shares1}
\end{figure}

If the scattering is due to on-site  or interstitial defects, the situation changes and we have two finite scattering processes in the effective  low-energy model: 
\begin{align}
    \hat{V}'_{S,1} &= \hat{\tau}_0 - \hat{\tau}_3 \label{eqn:low_scat_pot1} \\
    \hat{V}'_{S,2} &= \hat{\tau}_0 + \hat{\tau}_3 . \label{eqn:low_scat_pot2}
\end{align}
The overall impurity potential, which effectively renormalizes the order parameter, corresponds to a superposition of these. For random impurity distributions, the two scattering matrices are equally likely, and their average is ultimately given by the identity matrix~$\hat{\tau}_0$. This leads to scattering rates given by
\begin{align}
    \tau^{-1}_{\text{n}, i} &= \frac{\pi}{2} V_i^2 N(0) n S_i(n) \left(1+\braket{\hat{h}_{30}}^2\right),
\end{align}
and 
\begin{align}
    \tau^{-1}_{\text{sc}, i} &= \frac{\pi}{2} V_i^2 N(0) n S_i(n) \left(1-\braket{F_C^{\rm OP}}\right),
\end{align}
where $i = \{{\rm on}, {\rm is}\}$. Note that above the $\bm{k}$-dependence is implicit inside the brakets denoting the averages over the Fermi surface. Combining both equations leads to an effective scattering rate equal to
\begin{align}
    \tau_{\text{eff},i}^{-1} = \frac{\pi}{2} V_i^2 N(0) n S_i(n) \left( \braket{\hat{h}_{30}}^2  + \braket{F_C^{\rm OP}}  \right). \label{eqn:eff_scat_on&is}
\end{align}
The first term inside of the brackets is the same for all pairing symmetries. Thus, the difference between the effective scattering rate for different superconducting states is entirely governed by the average of the respective fitness function. As we can infer from Table~\ref{tab:Fc}, all gap functions have a finite fitness measure, whose size depends on the details of the microscopic Hamiltonian. This aspect was already emphasized in previous works \cite{Dentelski:2020,Cavanagh:2021,Sato:2020}. However, we also observe that the evolution of the share of symmetric configurations, $S_i(n)$, affects the scattering rate, which is a feature that does not depend on the structure of the Hamiltonian, but on the distribution of impurities. The influence of $S_i(n)$ on the renormalization of $T_c$ is illustrated in Fig.~\ref{fig:Tc_shares1}. 
As expected from the shape of $S_i(n)$ [Fig.~\ref{fig:shares_distr}], we obtain that all superconducting states are more sensitive to interstitial than on-site defects (assuming same magnitude of the scattering potential on the sites). Furthermore, we note that in case of strong impurity potentials, for which the superconducting state is suppressed by small impurity concentrations, there is an enhancement of the critical concentration by a maximum factor of approximately two (illustrated by Fig.~\ref{fig:Tc_shares1} b).

For intercalated configurations, the impurities can only occupy the first and fourth layer. Hence, the only finite scattering matrix in the low-energy model is given by $\hat{V}_{S,1}$~[Eq.~\eqref{eqn:low_scat_pot1}], which does not square to the identity matrix. This has important consequences for the scattering rates. First of all, since the gap functions not belonging to $A_{1g}$ commute (anti-commute) with $\hat{\tau}_0$ ($\hat{\tau}_3$), the superconducting scattering rate vanishes, i.e. $\tau^{-1}_{\rm sc, ic} = 0$. In contrast, the normal-state scattering rate is finite because all terms of the averaged Green's function fully commute with the scattering potential. The effective scattering rate is therefore equal~to
\begin{align}
    \tau_{\text{eff},\text{ic}}^{-1} = \tau_{\text{n},\text{ic}}^{-1} &= \pi V_{\rm ic}^2 N(0) n S_{\rm ic}(n) \left( 1 -  \braket{\hat{h}_{30} } \right)^2 \label{eqn:eff_scat_ic} .
\end{align}
We find that the normalized mass imbalance-term in the Hamiltonian, $\hat{h}_{30} $, is again an important factor that controls the  renormalization of the superconducting state in presence of impurities. However, in contrast to the cases of interstitial or on-site impurities, we obtain that a larger contribution of $\hat{h}_{30} $ to the normal-state Hamiltonian actually leads to a smaller effective scattering rate. Depending on the details of the normal-state Hamiltonian this could have profound consequences on the suppression of~$T_c$. 

In order to make better connections with experiments, we would like to emphasize here that the critical temperature is usually investigated as a function of the residual resistivity in the normal state~\cite{Kriener:2012,Smylie:2017}. The residual resistivity is proportional to the normal-state scattering rate, so it is sensible to study the evolution of the critical temperature not only with respect to the impurity concentration, but also with respect to the normal state scattering rate. We can write
\begin{align}
    \log \left(\frac{T_c}{T_{c0}}\right) = \Psi\left( \frac{1}{2} \right) - \Psi\left( \frac{1}{2} + \frac{1}{4\pi \tau_{\rm n} T_c} \frac{\tau_{\rm n}}{\tau_{\rm eff}} \right), \label{eqn:Tc_renorm2}
\end{align}
such that the ratio $\tau_{\rm n}/\tau_{\rm eff}$ can be thought of as the \emph{effectiveness} of the impurities in the superconducting state. If this ratio is large (small) the superconducting critical temperature is suppressed faster (slower) than naively expected for a single-band superconductor. In the case of on-site and interstitial scenarios, we find for superconducting states in non-trivial irreps
\begin{align}
   \frac{\tau_{\rm n}}{\tau_{\rm eff}}  = 
   \frac{\langle F_C^{OP} \rangle  + \langle\hat{h}_{30} \rangle^2}{1+\langle\hat{h}_{30} \rangle^2}.
\end{align}
From this ration, let us discuss two extreme limits. The first corresponds to $\braket{\hat{h}_{30}} \rightarrow 0$ such that $\tau_{\rm n}/\tau_{\rm eff} \rightarrow \langle F_C^{OP} \rangle $. In this limit, assuming OP orbitals we obtain $\braket{F_C^{OP}}  = \braket{ \hat{h}_{ab}^2} $ for a single $(a,b)$-term in the normal-state Hamiltonian (according to Table \ref{tab:Fc}). If this term is dominant, $\tau_{\rm n}/\tau_{\rm eff} \rightarrow 1$ and the superconductor behaves as expected for simple single-band scenarios. On the other hand, if the specific $(a,b)$ term is negligible, $\tau_{\rm n}/\tau_{\rm eff} \rightarrow 0$ and the superconducting state is suppressed at a much slower rate than expected from a naive estimation.
The second limit corresponds to $\braket{\hat{h}_{30} } \rightarrow 1$, with all other $\braket{\hat{h}_{ab}^2 } \rightarrow 0$. In this case, the ratio $\tau_{\rm n}/\tau_{\rm eff} \rightarrow 1 $ and the superconducting state is suppressed at the usual rate in terms of the normal state scattering rate. For intercalated impurities the ratio $\tau_{\rm n}/\tau_{\rm eff} = 1$, which is also in accordance with the expectation for simple superconductors. 

From this discussion, we can conclude that an arbitrarily robust unconventional superconducting state is possible in layered materials with OP orbitals stemming from distinct types of atoms under the condition that the $(a,b)$-terms enumerated in Table \ref{tab:Fc} (top) for each superconducting state are the least dominant terms in the normal-state Hamiltonian. Let us analyze next how these qualitative results change for EP basis states.

% case 2 %%%%%%%%%%%%%%%%%%%%%%%%%%%%%%%%%%%%%%%%%%%%%%%%
\subsubsection{EP orbitals from different atoms} 
According to Table~\ref{tab:tableI}, for basis states with EP, such as $\{P1_z^+, P2_z^+\}$, we obtain the same scattering matrices as in the previous subsection. This means that polar impurities play again no role in the renormalization of the superconducting state. The only differences compared to the OP case arise due to the changes in the structure of the Hamiltonian. Naturally, these differences are more of quantitative than of qualitative nature. 

For example, for on-site or interstitial distributions we find
\begin{align}
    \tau'^{-1}_{\text{n}, i} &= \frac{\pi}{2} V_i^2 N(0) n S_i(n) \left(1+\braket{\hat{h}_{30}}^2+\braket{\hat{h}_{10}}^2\right),
\end{align}
and 
\begin{align}
    \tau'^{-1}_{\text{sc}, i} &= \frac{\pi}{2} V_i^2 N(0) n S_i(n) \left(1-\braket{F_C^{\rm EP}}\right),
\end{align}
with $i = \{{\rm on}, {\rm is}\}$. Consequently, the effective scattering rate becomes
\begin{align}
    \tau_{\text{eff},i}'^{-1} &= \frac{\pi}{2} V_i^2 N(0) n S_i(n) \nonumber \\
    &\hspace{1cm}\times \left( \braket{\hat{h}_{30} }^2  + \braket{\hat{h}_{10} }^2  + \braket{F_C^{\rm EP}}  \right). \label{eqn:eff_ep_scat_on&is}
\end{align}
The only difference to Eq.~\eqref{eqn:eff_scat_on&is} is therefore a different prefactor determined by the microscopic structure of the Hamiltonian. Surprisingly, for intercalated impurities we obtain exactly the same results as before, since the $\hat{h}_{10} $-term of the Hamiltonian anti-commutes with $\hat{\tau}_3$. Consequently, its contribution vanishes in the normal-state scattering rate and we again end up with Eq.~\eqref{eqn:eff_scat_ic}. From this we can deduce that disorder located in the vdW~gap has the same effect in OP and EP two-orbital models stemming from distinct types of atoms. 

For the EP orbitals scenario, the discussion of the evolution of the critical temperature as a function of the normal-state scattering rate is similar to the one given above for the OP scenario. We can again conclude that an arbitrarily robust unconventional superconducting state is theoretically possible in layered materials with EP orbitals stemming from distinct types of atoms under the condition that the $(a,b)$-terms enumerated in Table \ref{tab:Fc} (bottom) are not the dominant terms in the normal-state Hamiltonian. Note that the number of terms in Table \ref{tab:Fc} is four, in contrast to two for the OP scenario. This means that superconductors stemming from Fermi surfaces formed by EP orbitals are generally much less likely to be robust than the ones stemming from OP orbitals. The condition for robustness would require four out of the six symmetry allowed terms in the normal-state Hamiltonian to be negligible.

% case 3 %%%%%%%%%%%%%%%%%%%%%%%%%%%%%%%%%%%%%%%%%%%%%%%%
\subsubsection{OP orbitals from same atoms} 
\begin{figure}[t!]
	\includegraphics[width=0.65\columnwidth]{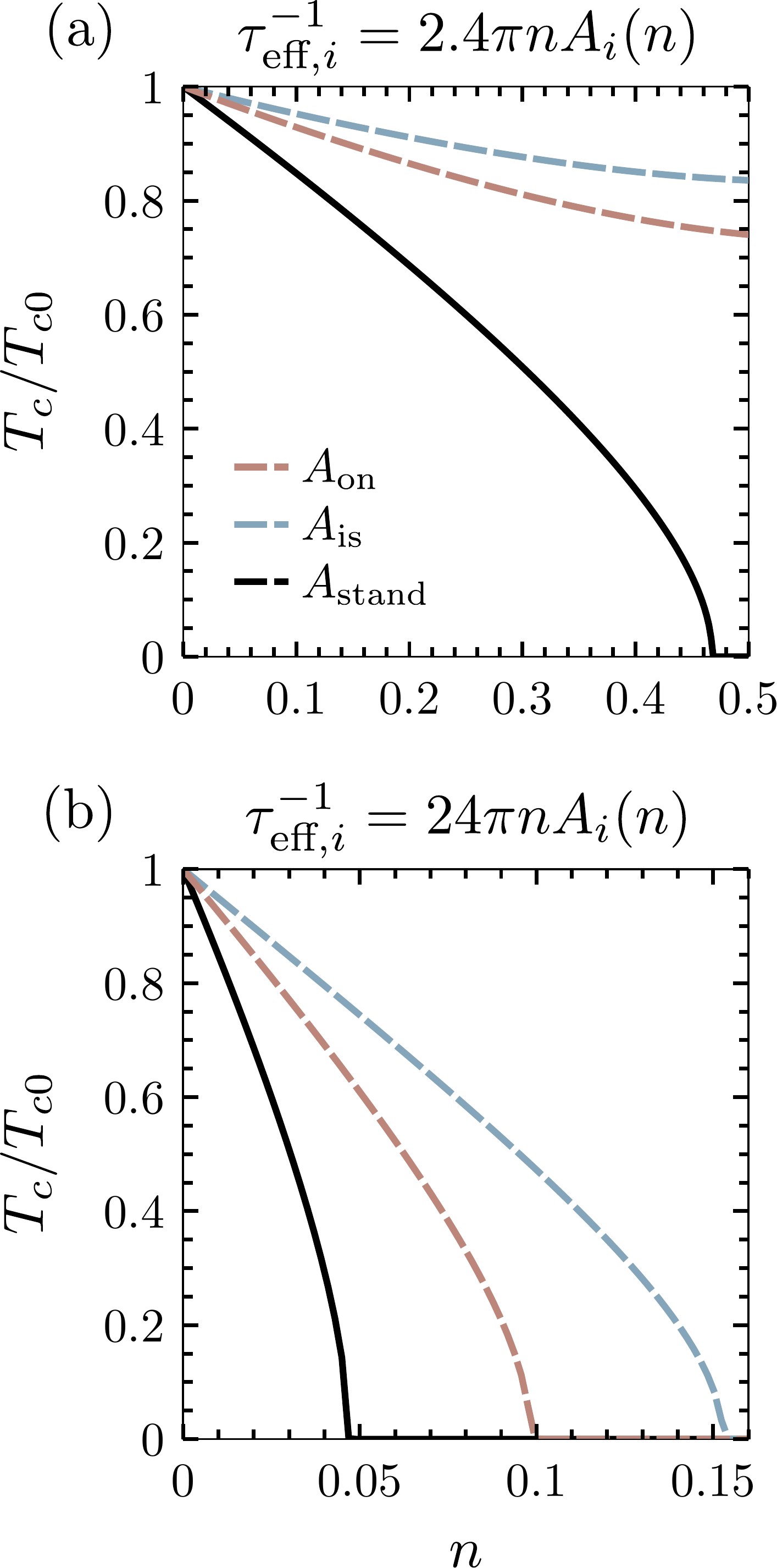}
	\caption{Supression of the normalized critical temperature $T_c/T_{c0}$ as a function of the impurity filling fraction $n$, if only the anti-symmetric share $A(n)$ of impurity distributions contributes, calculated from Eqs.~(\ref{eqn:Tc_renorm},~\ref{eqn:eff_diflay_scat_on&is_rewr}). We used the anti-symmetric shares $A_{\rm on/is}$ calculated in Sect.~\ref{sec:impurities}. We again choose a small and large prefactor to illustrate the effects of a weak and strong scattering potential, respectively.}
	\label{fig:Tc_shares2}
\end{figure}

If the states occupy the same type of atoms, as it is the case for $\{P2_z^+, P2_z^-\}$, the renormalization changes substantially. From Table \ref{tab:tableI}, we see that polar impurities, with scattering potential in the low-energy basis is given by $\hat{\tau}_1$, now suppress the superconducting state. The corresponding scattering rates read 
\begin{align}
    \tau''^{-1}_{\text{n}, \text{po}} &= \frac{\pi}{2} V_{\text{po}}^2 N(0) n A_{\text{po}}(n) \left(1-\braket{\hat{h}_{30}}^2\right),
\end{align}
and 
\begin{align}
    \tau''^{-1}_{\text{sc}, \text{po}} &= C_{\rm sc, po}^{ab}\frac{\pi}{2} V_{\text{po}}^2 N(0) n A_{\text{po}}(n) \left(1-\braket{F_C^{\rm OP}}\right),
\end{align}
where we obtain a different overall sign in the last equation due to the $C_{\rm sc, po}^{ab}$ factors [Table~\ref{tab:Cn}]. Subtracting the scattering rates leads now to 
\begin{align}
    \tau_{\text{eff}, \text{po}}''^{-1} &= \frac{\pi}{2} V_{\rm po}^2 N(0) n A_{\rm po}(n) \nonumber \\ 
    &\hspace{0.5cm}\times \left[
     1 - \braket{\hat{h}_{30} }^2 - C_{\rm sc, po}^{ab}  \left(1-\braket{F_C^{\rm OP}} \right) \right]. \label{eqn:eff_scat_po}
\end{align}
Note that again the scattering rates for different gap functions merely differ by the overall factor in the square brackets. Assuming the system is purely populated by polar impurities such that the share of anti-symmetric configurations stays constant, i.e. $A_{\rm po}(n)=1$, Eq.~\eqref{eqn:eff_scat_po} yields a renormalization which only depends on the details of the microscopic Hamiltonian. In Fig.~\ref{fig:Tc_shares2}, we show how this feature can influence the robustness of the superconducting state irrespective of the expansion coefficients of the normal-state Hamiltonian. 

For on-site, interstitial and intercalated impurities the scattering rates contain both symmetric and anti-symmetric shares. We obtain 
\begin{align}
    \tau''^{-1}_{\text{n}, i} &= \frac{\pi}{2} V_{i}^2 N(0) n \Big[ S_i(n) \left(1+ \braket{\hat{h}_{30} }^2\right) \nonumber \\
    &\hspace{2.5cm}+  A_{i}(n) \left( 1-\braket{\hat{h}_{30}}^2 \right) \Big],
\end{align}
and 
\begin{align}
    \tau''^{-1}_{\text{sc}, i} &= \frac{\pi}{2} V_{i}^2 N(0) n \Big[ S_i(n) \left(1- \braket{F_C^{\rm OP}}\right) \nonumber \\
    &\hspace{2.5cm}+ C_{\text{sc}, i}^{ab}  A_{i}(n) \left( 1-\braket{F_C^{\rm OP}} \right) \Big],
\end{align}
with $i=\{\text{on}, \text{is}, \text{ic}\}$. Note that not all non-A$_{1g}$ gap functions commute with $\hat{\tau}_1$, hence we have to include the different $C_{\text{sc}, i}^{ab}$ in the second line of the last equation. The effective scattering rate is 
\begin{align}
    \tau_{\text{eff}, i}''^{-1} &= \frac{\pi}{2} V_i^2 N(0) n  \Big\{1+\left[S_i(n) - A_i(n) \right] \braket{\hat{h}_{30} }^2  \nonumber \\
    &\hspace{1cm} -\left(1-\braket{F_C^{\rm OP}}\right) \left[  S_i(n) + C_{\text{sc}, i}^{ab} A_i(n) \right] \Big\}, \label{eqn:eff_diflay_scat_on&is}
\end{align}
where we used $S_i(n) + A_i(n) = 1$. 
We observe that if the states originate from the same layers, we cannot in general separate the symmetric and anti-symmetric shares from the microscopic details of the Hamiltonian. Similar to the previous cases, we find that the explicit structure of the shares influences the scattering rate. Analyzing to what extent, however, requires a more detailed calculation of the microscopic parameters, which is not the scope of this paper. Nevertheless, we can deduce from Eq.~\eqref{eqn:eff_diflay_scat_on&is} that there might be a fine-tuned value of parameters, for which only the anti-symmetric share contributes. Rewriting Eq.~\eqref{eqn:eff_diflay_scat_on&is}, 
\begin{align}
    \tau_{\text{eff}, i}''^{-1} &= \frac{\pi}{2} V_i^2 N(0) n  \Big\{ S_i(n) \left( \braket{\hat{h}_{30} }^2 + \braket{F_C^{\rm OP}} \right)  \nonumber \\
    &\hspace{0cm} + A_i(n) \left[ 1 - \braket{\hat{h}_{30} }^2 - \left(1-\braket{F_C^{\rm OP}}\right) C_{\text{sc}, i}^{ab}  \right] \Big\}, \label{eqn:eff_diflay_scat_on&is_rewr}
\end{align}
we see that this is the case if 
\begin{align}
    \braket{\hat{h}_{30} }^2 + \braket{F_C^{\rm OP}} = 0. 
\end{align}
Assuming $\braket{F_C^{\rm OP}}\rightarrow 0$ and $C_{\text{sc}, i}^{ab}=-1$, the renormalization of the critical temperature is governed by the anti-symmetric share of scattering potentials, as shown in Fig.~\ref{fig:Tc_shares2}. Intriguingly, we observe that the robustness of the superconducting state increases drastically for small scattering potential, since the share of anti-symmetric configurations approaches zero if the system is gradually filled with on-site or interstitial impurities. As an aside, we also have to emphasize that for intercalated impurities only our basis choice of $\{P2_z^+, P2_z^-\}$ leads to the expression given in Eq.~\eqref{eqn:eff_diflay_scat_on&is}. Naturally, choosing a basis which originates from the inner layers yields no renormalization for intercalated distributions.

Continuing with the discussion of the evolution of the critical temperature as a function of the normal-state scattering rate, we find now according to Eq. \eqref{eqn:Tc_renorm2},
\begin{widetext}
\begin{align}
   \frac{\tau_{\rm n}}{\tau_{\rm eff}}  = 
   \frac{1+\left[S_i(n) - A_i(n) \right] \braket{\hat{h}_{30} }^2  -\left(1-\braket{F_C^{\rm OP}}\right) \left[  S_i(n) + C_{\text{sc}, i}^{ab} A_i(n) \right]}{1+\left[S_i(n) - A_i(n) \right] \braket{\hat{h}_{30} }^2}
\end{align}
\end{widetext}
As before, we want to consider two extreme limits. The first concerns $\braket{\hat{h}_{30} } \rightarrow 0$, in which case the ratio simplifies~to
\begin{align}
   \frac{\tau_{\rm n}}{\tau_{\rm eff}}  = 
   {1 -\left(1-\braket{F_C^{\rm OP}}\right) \left[  S_i(n) + C_{\text{sc}, i}^{ab} A_i(n) \right]},
\end{align}
and becomes equal to $\braket{F_C^{\rm OP}}$ for $C_{\text{sc}, i}^{ab}= +1$. As the normalized fitness parameter satisfies $0<\braket{F_C^{\rm OP}}<1$, the ratio is necessarily smaller than one, and the superconducting state is generally less suppressed than in the naive single band scenario. For $C_{\text{sc}, i}^{ab}= -1$, the form of the ratio is not as simple, but the conclusion, based on the fact that $0\leq \tau_{\rm n}/\tau_{\rm eff}\leq 1$,  is the same. Overall, we find that for the scenario of two OP orbitals stemming from the same type of atoms, the conclusion that the superconducting state can be arbitrarily robust as long as the normal-state Hamiltonian terms $(a,b)$ contributing to $\braket{F_C^{\rm OP}}$ are negligible, is essentially the same as for the case of two OP orbitals stemming from distinct types of atoms.

%%%%%%%%%%%%%%%%%%%%%%%%%%%%%%%%%%%%%%%%%%%%%%%%%%%%%%%%%

\section{Conclusion}\label{sec:conclusion}

Inspired by the open questions concerning the robustness of the superconducting state found in materials belonging to the family of doped Bi$_2$Se$_3$, we have investigated  the effects of the impurity location and orbital content at the Fermi surface on the critical temperature of  layered superconductors. We started revisiting the microscopic description in the layer basis to faithfully account for four distinct impurity configurations: on-site, interstitial, intercalated and polar. After moving to the orbital basis by projecting our model into two low-lying degrees of freedom, we found three fundamentally distinct effective two-orbital bases. This allowed us to discuss whether the symmetric and anti-symmetric shares of the scattering potentials were active or inactive in each of these cases. In particular, we find that choosing two orbitals stemming from distinct types of atoms, the anti-symmetric part of the scattering potential is inactive, irrespective of the relative parity of the orbitals. We then elaborated on the renormalization of the superconducting critical temperature within the self-consistent Born approximation by providing closed-form expressions for the effective scattering rate. For simplicity, we restricted ourselves here to the unconventional states not belonging to the $A_{1g}$ representation. 
We obtain that the effective scattering rate depends on three important properties: i) the commutation/anti-commutation relation between the normal-state Hamiltonian and the scattering potential;  ii) the commutation/anti-commutation relation between the superconducting order parameter and the scattering potential; and iii) the commutation/anti-commutation relation between the normal-state Hamiltonian and the superconducting order parameter (known as the superconducting fitness measure). 

For the main case of interest, OP orbitals stemming from distinct types of atoms, we observe that purely polar impurities do not suppress the superconducting state. Moreover, on-site and interstitital impurities behave qualitatively the same: the superconducting state is more robust the smaller the $(3,0)$-term in the normal-state Hamiltonian. Nevertheless, on-site impurities are generally less effective in suppressing the superconducting state compared to interstitial impurities (assuming the magnitude of the scattering potential and the normal-state Hamiltonian are the same). This is due to the distinct evolution of the symmetric share as a function of doping. The superconducting state is also possibly more robust than usual for intercalated impurities, but now the robustness is enhanced if the contribution of the $(3,0)$-term to the normal-state Hamiltonian increases. 

We also conclude that superconductors emerging from Fermi surfaces formed by two OP orbitals are generally more robust than superconductors stemming from Fermi surfaces with EP orbitals. This effect is purely controlled by the fitness measure $\braket{F_C}$. Furthermore, considering two OP orbitals stemming from the same type of atoms as the basis states, we conclude that these are affected by purely polar impurities, in contrast to the scenario where the orbitals stem from different types of atoms.

In addition, within the scenario of OP orbitals, we discovered a new mechanism to enhance the robustness of superconducting states. In case the anti-symmetric share of the impurity potential is dominant, the superconducting state can also be exceptionally robust to on-site, interstitial and intercalated impurities since the anti-symmetric share decreases with doping.

Finally, we highlighted the difference in discussing the robustness of the superconducting state as a function of the impurity concentration or as a function of the normal-state scattering rate. Our findings show that an enhanced robustness can be observed for both on-site and interstitial impurities when analyzing the critical temperature either as a function of impurity concentration or $\tau_{\rm n}^{-1}$. On the other hand, an enhanced robustness for intercalated impurities only appears when analyzing the evolution of the superconducting critical temperature with the impurity concentration. These results can have important implications for the interpretation of experiments and the identification of the impurity location in layered materials.

\section*{Acknowledgements}
We would like to thank Manfred Sigrist, Philip Brydon, and David C. Cavanagh for helpful discussions. A.R. acknowledges the financial support of the Swiss National Science Foundation (SNSF) through an Ambizione Grant No.~186043 and B.Z. the financial support of the SNSF through Division II (No. 184739).

\bibliography{draft}{}

\end{document}